\documentclass[a4paper,conference]{IEEEtran}
\IEEEoverridecommandlockouts
\usepackage[left=1.57cm,right=1.57cm,top=2.4cm,bottom=4cm]{geometry}
\usepackage{cite}
\usepackage{amsmath,amssymb,amsfonts}
\usepackage{algorithmic}
\usepackage{graphicx}
\usepackage{textcomp}
\usepackage{xcolor}
\usepackage{url}  
\usepackage{multirow}
\usepackage[sc]{mathpazo}

\usepackage{graphicx}
\usepackage{multirow}
\usepackage{times}  
\usepackage{helvet}  
\usepackage{courier}  
\usepackage{url}  
\usepackage{graphicx}  
\usepackage{enumitem} 
\setlist[itemize]{noitemsep} 

\usepackage{array}
\newcolumntype{H}{>{\setbox0=\hbox\bgroup}c<{\egroup}@{}}

\usepackage{balance}
\usepackage{comment}
\usepackage{xcolor}
\usepackage{soul}
\usepackage[utf8]{inputenc}
\usepackage[small]{caption}
\usepackage{amsfonts}
\usepackage{amsmath}
\usepackage{epsfig}
\usepackage{amsopn}
\usepackage{float}
\usepackage{epstopdf}

\usepackage{booktabs} 
\usepackage{algorithmic}
\usepackage{colortbl}
\usepackage[ruled]{algorithm2e} 

\usepackage{caption}
\usepackage{subcaption}
\usepackage{multirow}
\usepackage{tabularx}
\usepackage{bm}
\usepackage{rotating}
\usepackage{mathrsfs}
\usepackage[font=small]{caption}

\def\BibTeX{{\rm B\kern-.05em{\sc i\kern-.025em b}\kern-.08em
    T\kern-.1667em\lower.7ex\hbox{E}\kern-.125emX}}

\begin{document}

\title{ECG-Based Heart Arrhythmia Diagnosis Through Attentional Convolutional Neural Networks} 

\author{\IEEEauthorblockN{Ziyu Liu}
\IEEEauthorblockA{
\textit{University of New South Wales}\\
Sydney, Australia \\
ziyuliu813@gmail.com}
\and
\IEEEauthorblockN{Xiang Zhang}
\IEEEauthorblockA{
\textit{Harvard University}\\
Boston, USA \\
xiang\_zhang@hms.harvard.edu}
}

\maketitle

\begin{abstract}
Electrocardiography (ECG) signal is a highly applied measurement of the heart electrical activity for individual heart function and condition. Among the heart diseases can be indicate by ECG monitoring, much effort have been endeavored towards automatic heart arrhythmia diagnosis based on machine learning methods. 
However, traditional machine learning models require time-consuming preprocessing and laborious feature extraction which highly depend on domain knowledge. 
Here, we propose a novel deep learning model, 
named Attention-Based Convolutional Neural Networks (ABCNN) that taking advantage of CNN and multi-head attention, 
to work straight on raw ECG data and extract the informative dependencies automatically for accurate arrhythmia detection. The proposed method is evaluated by extensive experiments over a public benchmark ECG dataset. 
Our main task is to find the arrhythmia from normal heartbeats and, at the meantime, accurately recognize the heart diseases from five arrhythmia types. We also provide convergence analysis of ABCNN and intuitively show the meaningfulness of extracted representation through visualization.
%
The experimental results show that the proposed ABCNN outperforms the widely used baselines, which puts one step closer to intelligent heart disease diagnosis system.
\end{abstract}


\section{Introduction}

Electrocardiography (ECG) refers to a tracing of heart bio signal over a specific period of time, is commonly understand as the monitoring of heart beat and collected by the electrodes placing on ones chest~\cite{soman2005classification, chauhan2015anomaly, hou2019lstm}. The significant electrical variations of each heartbeat are measured by these electrodes based on the electrophysiological pattern of heart muscles after depolarizing and repolarizing. The typical pattern of a normal ECG record is known as PQRST complex waves, each signal wave generate by electrical activities from the  different conduction path within distinct part of heart. ECG signal is rich of critical information including the basic situation about the heart activities, 
as well as the size and position of heart chambers, thus, it is an highly applied noninvasive cardiac diagnostic procedure which used for a variety of scenarios, such as diagnosis of heart disease, congestive cardiac failure detection \cite{hossen2008identification}, heart rate monitoring \cite{schumann2002potential}, remote medicine service \cite{hsieh2012cloud}, homecare tracking \cite{lobodzinski2010integrated} as well as hearth care applications integrated in mobile devices \cite{mazomenos2013low}. 




In the heart diseases discovered and identified through the test of ECG, any disturbance within the heart's electrical system could lead to heart arrhythmia \cite{doi:10.1001/jama.1989.03420030127049}.
The symptoms commonly demonstrated as abnormal heart rate or irregular heartbeat, which acting as racing or slow heartbeat, may also result in chest pain and shortness of breath, dizziness and fainting.
Heart arrhythmia is usually divided into two types,
the first group of arrhythmia (e.g., tachycardia, bradycardia and ventricular fibrillation) require professional treatment without delay because the level of severe are fatal, in some cases, the invasive medical devices such as artificial pacemaker and cardiac defibrillator are needed when deal with more serious syndrome.
Other type of heart arrhythmia are not immediately life-threatening, nevertheless, proper treatments still should not be neglected. Thus, accurate and timely heart arrhythmia diagnosis is crucial.




However, the ECG-based diagnosis of heart arrhythmia faces several challenges.
To begin with, the existing studies depend on features (such as RR intervals~\cite{sadhukhan2012r}) of ECG signals that are manually refined~\cite{shinde2011comparison}.
The pre-processing of raw ECG signal requires large investment of time and effort, while the feature engineering could also be laborious owing to the artificial selection of the feature which heavily relies on domain knowledge.  
%
For example, QRS complex detection is used as a common approach to discover the shape of QRS wave from a whole cardiac cycle. This process requires manual operation and would fail to capture the information that is not covered by the QRS wave. 
In order to overcome the aforementioned limitations, we propose a novel deep learning model which can learn from raw ECG data and require little pre-processing and input-feature engineering.

Recently, a growing body of literature has illustrated the success of deep learning in areas like natural language processing~\cite{socher2012deep}, computer version~\cite{huang2015face}, as well as Brain-computer Interface (BCI)~\cite{deng2014deep}.
Among the typical deep learning methods, Convolutional Neural Networks (CNNs) have been widely applied to digital signal processing and obtained great achievements. 
In traditional CNN, input neurons are all weighted equally. However, in terms of ECG data, diverse heartbeat cycles have different importance.
For instance, the QRS wave lasts only for short period (around $ 0.08 \sim 0.12$ seconds) but occupies a considerable part of the information.

\begin{table*}[htb]
\centering
\caption{Redefine the MIT-BIH heart arrhythmia dataset ECG classes based on the AAMI standard}
\label{MIT-to-AAMI}
\resizebox{0.9\textwidth}{!}{%
\begin{tabular}{c|ccccc}
\hline
\begin{tabular}[c]{@{}c@{}}AAMI ECG \\ classes\end{tabular}       & N        & S  & V      & F    & Q    \\ \hline
Description      & \begin{tabular}[c]{@{}c@{}}Normal heartbeat that not included \\ in S,V,F or Q classes\end{tabular} & \begin{tabular}[c]{@{}c@{}}Supraventricular \\ ectopic beat\end{tabular}    & Ventricular ectopic beat       & Fusion beat       & Unknown beat          \\ \hline
\multirow{5}{*}{\begin{tabular}[c]{@{}c@{}}MIT-BIH \\ database\\ ECG classes\end{tabular}} & Normal beat (N)       & \begin{tabular}[c]{@{}c@{}}Atrial premature \\ beat (A)\end{tabular}        & \begin{tabular}[c]{@{}c@{}}Premature ventricular \\ contraction (V)\end{tabular} & \begin{tabular}[c]{@{}c@{}}Fusion of ventricular \\ and normal beat (F)\end{tabular} & Paced beat (/)        \\
& \begin{tabular}[c]{@{}c@{}}Left bundle branch \\ block beat (L)\end{tabular}      & \begin{tabular}[c]{@{}c@{}}Aberrated atrial \\ premature beat (a)\end{tabular}     & \begin{tabular}[c]{@{}c@{}}Ventricular escape \\ beat (E)\end{tabular}    &      & \begin{tabular}[c]{@{}c@{}}Fusion of paced \\ and normal beat (f)\end{tabular} \\
& \begin{tabular}[c]{@{}c@{}}Right bundle branch \\ block beat (R)\end{tabular}     & \begin{tabular}[c]{@{}c@{}}Nodal (junctional) \\ premature beat (J)\end{tabular}   &  &      & \begin{tabular}[c]{@{}c@{}}Unclassifiable \\ beat (Q)\end{tabular}      \\
& Atrial escape beat (e)          & \begin{tabular}[c]{@{}c@{}}Supraventricular premature \\ or ectopic beat (S)\end{tabular} &  &      &      \\
& \begin{tabular}[c]{@{}c@{}}Nodal (junctional)\\ escape beat (j)\end{tabular}      &    &  &      &      \\ \hline
\end{tabular}%
}
\vspace{-5mm}
\end{table*}

To address this issue, inspired by the success of attention mechanism~\cite{10.1145/3264959}, we propose Attention Based Convolutional Neural Network (ABCNN) with an attention layer that constraints the model to assign higher weight for the more informative signals.
The proposed ABCNN can automatically capture the distinctive features from the raw ECG signals for the heart arrhythmia detection. 

The attention mechanism is designed to pay more attention to the most informative part of the raw data such as QRS complex. Differ from the traditional QRS complex detection method, the attention weights in ABCNN is automatically learned by the model based on the training data. We summary our contributions as follow:
\begin{itemize}
	\item We propose a novel approach, ABCNN, to automate extract the distinctive representations that are directly learned from raw ECG signals, for precise recognition of heart arrhythmia. The proposed ABCNN, combining attention mechanism and CNN, can pay more attention to the important information in ECG signals.
	\item The proposed ABCNN is evaluated over a well-known heart arrhythmia dataset. The experimental results demonstrate that ABCNN outperforms a wide range of competitive state-of-the-art baselines.
\end{itemize}

Please find the dataset and code of model implementation at \url{https://github.com/ziyuliu-lion/heart-arrhythmia-diagnosis-with-deep-learning}.

\section{Preliminary knowledge and related work}
\label{ch:background}

\subsection{Preliminary knowledge of heart arrhythmia}

Heart arrhythmia happens when the heartbeats can not work correctly which will cause abnormal electrical impulses of the heart.
The too fast/slow or irregular heartbeat could directly lead to fault in the circulatory system that pumping blood to support your organs to functioning well. Atrial fibrillation (AFib) is the most common type of heart arrhythmia that can cause stroke and heart failure. 
There are 33.5 million people are diagnosed with AFib
worldwide~\cite{chugh2014worldwide}.
Therefore, the detection and diagnosis of heart arrhythmia either common or fatal are all worth of attention.

Following the ANSI/AAMI (Association for the Advancement of Medical Instrumentation) EC57:1998 standard which is used to detect cardiac rhythm disturbances and ST segment measurement,  
heart beats are commonly divided in to five typical categories including class N (normal), S (supraventricular), V (ventricular), F (fusion) and Q (question/unknown). The description and their corresponding categories in a benchmark (MIT-BIH~\cite{goldberger2000physiobank}) dataset are presented in table~\ref{MIT-to-AAMI}.

\subsection{Traditional ECG classification} 
\label{sec:ecg_classification}
A substantial amount of various studies on heart disease classification and diagnosis by adopting
traditional machine learning methods have been developed in past decades~\cite{osowski2004support, mahajan2017cardiac, soman2005classification, dalvi2016heartbeat, mohebbi2008detection}. These researches aim to design and apply a machine learning approaches (such as SVM~\cite{osowski2004support}, random forest~\cite{mahajan2017cardiac}, Naïve Bayes~\cite{soman2005classification}, shallow neural network~\cite{dalvi2016heartbeat}) in order to assist surgeons/cardiologist in the field for a more comprehensive analysis and diagnosis. For example,
Jha et al.\cite{jha2021tunable} proposes a tunable Q-wavelet transform to efficiently compress the ECG signals and fed the compact data to SVM to preform the cardiac arrhythmia patterns recognition.

However, there are several issues of traditional machine learning approaches that can not be disregarded.
To start with, the tedious and chaotic pre-processing of the raw ECG data is labour-intensive and time consuming, although it is essential to the traditional methods, the effect of it to the final performance is yet to see until the final step.
Furthermore, feature engineering (such as wavelet transform and time-domain/frequency-domain feature extractions) is also laborious and domain-dependent tasks. Not to mention the manually extracted features may not contribute positively to the performance of the optimal classification system.

\subsection{Deep learning-based ECG classification} 
\label{sec:deep_learning_based_ecg_classifier}
The classifiers established on top of deep neural networks are also an intense research trend~\cite{chauhan2015anomaly,yildirim2018novel,al2016deep,rajpurkar2017cardiologist, sakib2021rigorous}. 
For instance, Kiranyaz et al.~\cite{kiranyaz2016real} propose a 1-D CNN  classification system for patient-specific ECG monitoring which only takes a relatively small amount of common and individual data as input. This learning framework thus can integrate the  process of feature extraction and classification together to further enhance the classification results.  
In \cite{acharya2017automated}, Acharya et al. support diagnosis with a designed CNN model that can automatically learn differential features from ECG signals. The CNN model has 11 convolutional and pooling layers. The learned features are then fed into a fully-connected layer for diagnosis.

In summary, most of existing deep learning models for ECG signal analysis assign the same weights across different components in ECG signals, which prevent the model to focus on the real distinctive pattern. To this end, we propose ABCNN to assign higher weights for the more informative time slices.

\begin{figure*}[!h]
\centering
  \includegraphics[width=\linewidth]{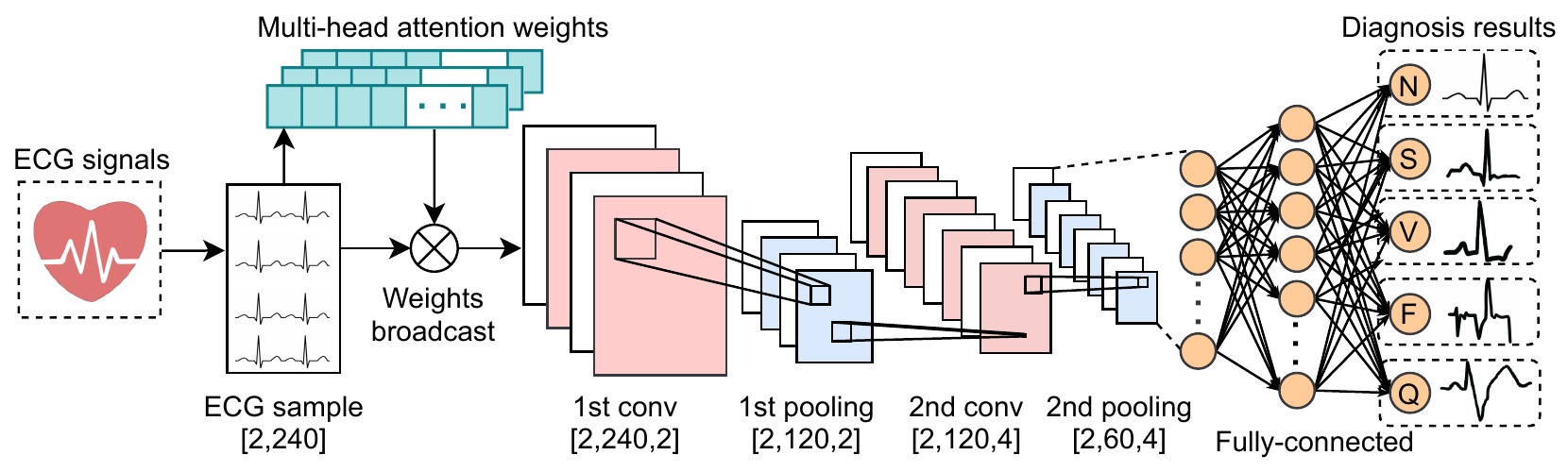} 
  \caption{The workflow of the proposed ABCNN. The model learns an attention weight for each time slice of the input EEG sample, in order to pay more attention to the important signatures. The weights are broadcast to the same shape with sample and update the sample through element-wise multiplication. The five nodes in the last layer produces the probability that how likely the input ECG sample associates with a certain heart arrhythmia type.
   }
  \label{fig:methodology}
\end{figure*}

\section{Methodology} 
\label{cha:methodology}

\subsection{Overview and motivation}
CNN has obtained great success in many research topics, such as computer vision~\cite{huang2015face} and recommendation system \cite{kiranyaz2016real}, due to the excellent high-level feature learning ability.
Although Recurrent Neural Networks (RNN; such as LSTM) are designed for temporal feature capturing, however, they are challenged by dealing with long dependencies and computational expensive for long sequence: our ECG signals are long sequences, so that we design our model on top of CNN base.
In general, a CNN architecture is composed of three essentially layers: convolutional layer, pooling layer, and fully connected layer. 
In this work, we propose ABCNN model that can capture the interactions between different EEG electrodes.
For better understanding, we take a real ECG sample from MIT-BIH dataset, a benchmark ECG collection, as an example to present our model.
The ECG signals are acquired from 2 leads while the sampling frequency is 360 Hz.  
For each heartbeat, we choose a time window of 0.76s including 240 time points and set the R peak locates at the central of time window.
We regard each time window as a segment that associates with a label indicating patient's status. In this way, the ECG signals in a single segment can be denoted by a matrix (with shape $[2, 240]$) where
each row represents a ECG channel while each column denotes a sampling point.

Standard CNN architecture treats all input neurons equally, nevertheless, different heartbeat periods take different importance naturally.
For instance, the QRS complex in ECG sample, which only lasts for $0.08-0.12$ seconds (around 10\% length in a heartbeat) but contribute the most crucial message. Thus, the model should pay more attention to the QRS component. 
Addressing this problem, we develop ABCNN which assign the informative time points higher importance through attention mechanism. In particular, we involve a well-designed attention layer to encourage the model to pay more attention to the more significant time slices.

\subsection{Attention-Based Convolutional Neural Networks} 
\label{sec:attention_based_convolu}
As shown in Figure~\ref{fig:methodology}, the proposed ABCNN model is stacked by: a input layer that receives the raw ECG signals, an attention layer that measures the importance of input sample and controls how much attention the model should pay to each signal slice, the 1st convolutional and the 1st pooling layer, the 2nd convolutional and the 2nd pooling layer, the 1st and the 2nd fully connected layer, and the output layer which further generates labels. 
Next we explain every layer of ABCNN in detail. The convolutional layer contains multiple random-initialized kernels to operate convolutional transformation over the input signals~\cite{zhang2020survey}. 
Then, we adopt a pooling layer to reduce the dimension of embeddings and also capture necessary spatial features.
In general, a pooling layer is applied after a convolutional layer, for the aiming of mitigating the computational expenses. In addition, adding a pooling layer reduces the  number of demanded trainable parameters and benefit to prevent overfitting. At last, a fully connected layer, a.k.a. dense layer, is composited of multiple independent neurons. Every neuron in a fully connected layer is connected with all neurons in the successive layer (that's why it's called \textit{fully-connected layer}). However, there's no connections among the same layer. 
Next, we introduce the notation and model structure.





\subsubsection{Notation}
We describe the working principles of ABCNN at a single heartbeat (e.g., sample) level and omit the sample index for simplicity.
We denote the embeddings in the $i$-th layer ($i=1,2, \cdots, 8$) by $X_i \in \mathbb{R}^{[n_i, K_{i},d_i]}$, where $n_i$, $K_i$ and $d_i$ denote the number of channels, dimension of representation, and the representation depth of the $i$-th layer, respectively. In the input layer, a single ECG sample is denoted by $X_1 \in \mathbb{R}^{[2, 240, 1]}$. In the following introduction, we use exact numbers to denote the shape of data for better understanding.


\subsubsection{Multi-head attention}
We design ABCNN to pay more attention to the important ECG peaks by integrating attention mechanism. In specific, we assign a weight to each time slice in the input ECG sample. 
For the input segment with 240 time slices, we use a nonlinear fully-connected layer to learn an attention weight vector with 240 elements, where each element corresponds to the importance of a certain time slice. A softmax layer is used to normalized all the elements to make the sum of them equals to 1. We then broadcast the weights to the same shape as $X_i$ and conduct element-wide multiplication to adjust the scale of each dimension. Furthermore, to increase the express ability, we introduce multi-head attention to independently calculate multiple attention weights~\cite{vaswani2017attention}. Then, we take the average of all attention heads' results as the final attention weights.

\subsubsection{ABCNN structure}
We set both the kernal size and the stride in the first convolutional layer as $[1,1]$. The stride means the sliding distance of kernel along with x-axis and y-axis during the convolution operation. 
Moreover, we set the padding format as zero-padding and keep the shape of features unchanged after convolution. 
As we use two kernels in the first convolutional layer, the depth of features increased from 1 to 2, and the shape of $X_2$ becomes $[2,240,2]$.  A pooling layer follows the convolutional layer. After each pooling layer, we use ReLU as the activation function to process the results of convolutional operation. In detail, we set the pooling window as $[1,2]$ and the sliding stride as $[1,1]$ in the first pooling layer. We adopt max pooling which means that we take the maximum value in each pooling window~\cite{nagi2011max}.
By going through the pooling layer, the representation's shape shrinks according to the pooling parameters. Thus, $X_3$ has a shape of $[2,120,2]$. Similarly, we use 4 kernels in the second convolutional layer while the kernel size is set as $[1,2]$ and both the x-axis and y-axis strides are 1. As a result, the output of this layer has shape of $[2,120,4]$. In the second pooling layer, we select the same pooling window and stride as the previous pooling layer: resulting to the shape of representation as $[2, 60,4]$. 


At last, we flat the learned discriminative features (which is a 3-D tensor) into a 1-D vector and then feed it into a classifier that is composed of two fully-connected layers. For example, we unfold $X_5$ from size $[2, 60,4]$ to $X_6$ with size of $[1,480]$. As we are making 5-class classification, the output layer contains 5 neurons. The predicted probabilities $X_8$ is a vector with 5 elements where each element corresponding to the probability of a particular label.
We adopt binary cross-entropy to calculate the loss function and optimize the loss with widely-used Adam optimizer.

%

\section{Experimental setting} 
\label{sec:experimental_setting}

\subsection{Dataset}
Our model is evaluated over a public benchmark, MIT-BIH heart Arrhythmia dataset\footnote{\url{https://physionet.org/content/mitdb/1.0.0/}} \cite{goldberger2000physiobank}, that has been widely used in diagnosis of heart arrhythmia.
This dataset contains 4,000 long-term Holter recordings that last for 24 hours, collected from 47 patients (60\% inpatients and 40\% outpatients; 25 males aged from 32 to 89 yers; 22 female aged from 23-89 years).
In the preprocessing stage, 48 adequate quality recordings are selected and each of them last for about 30 mins. 
We focus on two subsets of the dataset: the first one has 23 recordings representing typical heart beat signals; the second one covers the clinically significant but less common arrhythmias (such as junctional, supraventricular arrhythmias, complex ventricular, and conduction abnormalities).
The dataset measures 2-lead ECG signals (MLII and V5) with 360 Hz sampling rate.
Every heartbeat last for about 1 second, thus, we select 240 sampling slices (around 0.76 second) of each beat as input segment to detect arrhythmia. While selecting the segment, we select 240 sampling slices prior the R peak and 240 sampling slices after the R peak. 


\subsection{Multi-class arrhythmia diagnosis scenario}
According to the characteristics of heart arrhythmia, we split the heartbeats from MIT-BIH dataset into 5 typical classes.
All the 48 subjects, including the subset with rare symptoms, are used for automatic heart arrhythmia classification.
We select 2,000 samples from each subject and get 96,000 samples in total. Each sample has 2 channels while each channel has 240 successive observations. We randomly split the dataset into training (80\%) and testing (20\%) set. 
In a single sample, we set the R peak at the middle of the time window.
The standard (z-score) normalization is applied to the raw ECG data to prevent the influence of channels' varying scales. 

\begin{figure*}[ht]
    \centering
    \begin{subfigure}[t]{0.4\textwidth}
        \centering
        \includegraphics[width=\textwidth]{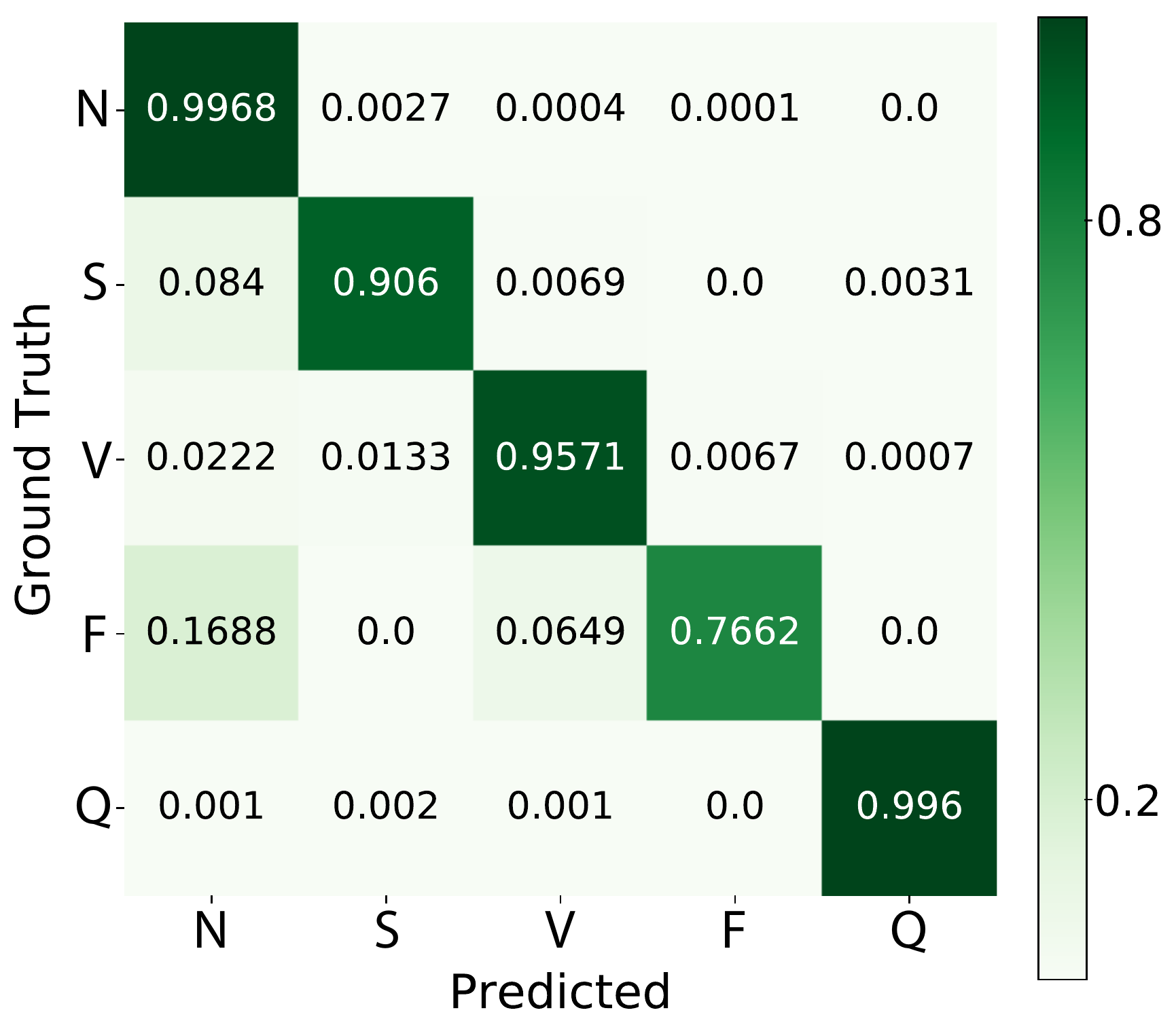}
        \caption{Confusion Matrix}
    \end{subfigure}%
    \hspace{2cm}
    \begin{subfigure}[t]{0.45\textwidth}
        \centering
        \includegraphics[width=\textwidth]{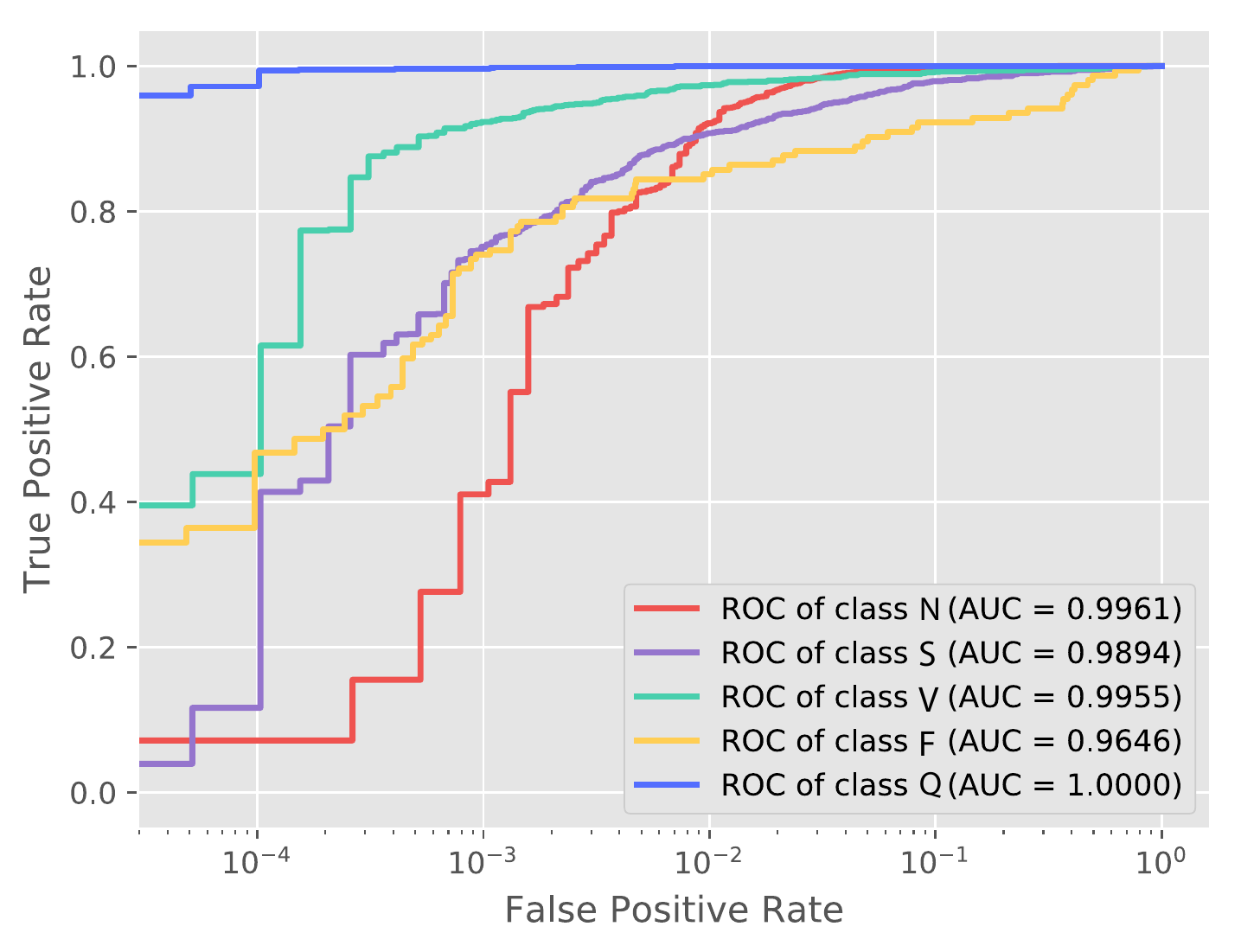}
        \caption{ROC with AUC}
    \end{subfigure}%
    \caption{Confusion matrix and ROC curves
    }
    \label{fig:ROC_5}
\end{figure*}

\subsection{Evaluation metrics}
The number of samples associated with different arrhythmia types is highly imbalance, thus we use Receiver Operating Characteristic (ROC) curves and Area Under the Curves (AUC) score to evaluate the model's performance as they not only consider the predicted results but also consider the predictive probabilities (confidence level on the predicted label). 
The AUC score falls in the range of $[0.5,1]$, the higher the better. 
Although the commonly used evaluation metrics (such as accuracy, precision, recall, and F-1 score) are not suitable as they all sensitive the sample amount, we use them to show the model's performance on each single class.

\begin{table}[]
\centering
\caption{The comparison among our method ABCNN and widely used baseline in multiclass arrhythmia detection.}
\label{tab:multi_comparing}
\begin{tabular}{llcH@{\hspace*{-\tabcolsep}}}
\hline
Literature & Model & AUC & ACC \\ \hline
\cite{barrella2014identifying} & SVM & 0.9514 $\pm$ 0.005 & 0.9365 \\
\cite{mahajan2017cardiac} & RF & 0.9704 $\pm$ 0.004 & 0.961 \\
\cite{saini2015classification} & KNN & 0.9126 $\pm$ 0.009 & 0.9535 \\
\cite{acharya2017automated} & CNN & 0.9655 $\pm$ 0.025 & 0.8368 \\
\cite{hou2019lstm} & LSTM & 0.6297 $\pm$ 0.034 & 0.8116 \\
Ours & ABCNN & \textbf{0.9896 $\pm$ 0.010} & 0.8634 \\ \hline
\end{tabular}
\end{table}

\begin{table}[]
\centering
\caption{Summary of diagnosis results in each specific arrhythmia category. } 
\label{tab:multicalss_report}
\begin{tabular}{c|cccc}
\hline
\textbf{Label} & \textbf{Precision} & \textbf{Recall} & \textbf{F-1} & \textbf{Support} \\
\hline
\textbf{N} & 0.9902 & 0.9968 & 0.9965 & 16884 \\
\textbf{S} & 0.9476 & 0.9061 & 0.9263 & 1298 \\
\textbf{V} & 0.9795 & 0.9571 & 0.9682 & 1351 \\
\textbf{F} & 0.9147 & 0.7662 & 0.8339 & 154 \\
\textbf{Q} & 0.995 & 0.996 & 0.9955 & 993 \\
\textbf{Average} & 0.9865 & 0.9868 & 0.9865 & 20680 \\ \hline
\end{tabular}
\end{table}

\subsection{Hyper-parameter tuning} 
We conduct preliminary experiments to tune hyper-parameter, here we report the settings: the first convolutional layer has 4 kernels with size $[2,2]$ and stride of $[1,1]$; the first pooling layer has the kernel size as $[1,2]$ and strides of $[1,1]$; the second convolutional layer has 8 kernels which have the identical kernel and stride size with the first convolutional layer; the second pooling layer has the kernel size as $[1,2]$ and strides of $[1,1]$.
The two fully connected layers have 240 and 60 hidden nodes, respectively. The output layer has 5 nodes where each node corresponds to an ECG classes. The number of heads for attention mechanism is 32. We use Adam optimizer with a learning rate of 0.0005. We use early-stopping strategy that stop the iteration if the testing loss not decrease over the prior consequent 20 epochs. A dropout layer with 0.3 dropout rate is added to the first fully-connected layer. 
The experiment settings (e.g., training set and testing set splitting, learning rate, number of epochs), without specific clarification, are the same for all the methods.

\subsection{Baselines}
We compare the proposed ABCNN with a number of competitive typical machine learning baselines including Support Vector Machine (SVM), Random Forest (RF), K-nearest neighbors (KNN). 
We report the key parameters of baselines: KNN with 3 nearest neighbors; SVM with RBF kernel; RF with 50 trees. To further demonstrate the effectiveness of ABCNN, we compare with the standard CNN and long short-term memory (LSTM), the latter is another very popular deep learning architecture. For fair comparison, the CNN baseline has the same parameters as ABCNN (introduced in previous subsection).
except attention module. For the baselines with LSTM model, we reimplement the model with 2 LSTM layers where each layer has 6 cells; the time step is 240. More experimental settings are presneted in our public code.

\section{Results} 
\label{cha:experiments}


\subsection{Classification results} 
\label{sec:classification_results}

The comparison results of ABCNN and several competitive baselines are shown in Table~\ref{tab:multi_comparing}. We can observe that ABCNN obtains the best performance by the AUC of 98.96\%, which outperforms all the well-known traditional machine learning-based or deep learning-based classifiers. For a fair comparison, we use AUC to evaluate the results as the dataset are imbalanced. We notice that LSTM has relative poor AUC which may caused by the long sequence of ECG samples.
For more details in each single heart arrhythmia type, the classification report in Table~\ref{tab:multicalss_report} show the classification metrics such as recall, precision, F-1 score, and support for every specific class. The `support' denotes the number of samples in test set that associating with the particular class. 
%
%
In addition, we report the confusion matrix, ROC curves, and AUC scores for each class (Figure~\ref{fig:ROC_5}). 
The X-axis of ROC figure is set in log scaled for better visualization.
The class F has the lowest accuracy and AUC score, indicating the fusion beats is an arrhythmia type that is most difficult recognized. 
Additionally, we can learn from the confusion matrix that a small ratio of misclassified fusion beats (16.88\%) are classified as normal beats (i.e., class 0). This result is reasonable because the fusion beats combines normal beats and ventricular beats. The class N and class Q obtain the highest performances
as they contain the highest and the second highest number of samples in the datasets. Moreover, although the training time is above 20 mins, the inference time of our model is less than 0.1s, which is sufficient in real-world deployment.

\begin{figure}[]
    \centering
    \begin{subfigure}[t]{0.5\textwidth}
        \centering
        \includegraphics[width=0.8\textwidth]{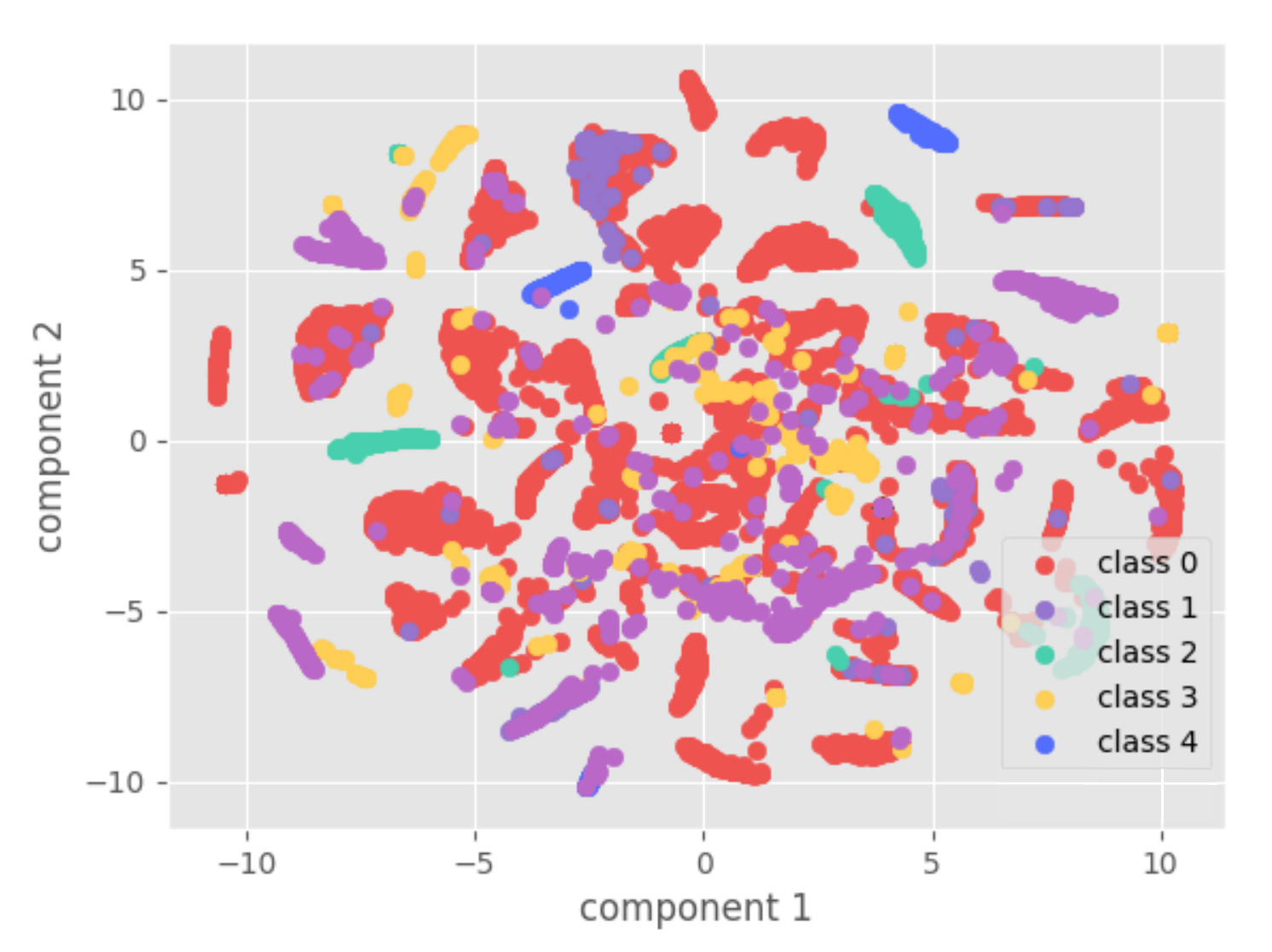}
        \caption{Raw data}
    \end{subfigure}%
    \\
    \begin{subfigure}[t]{0.46\textwidth}
        \centering
        \includegraphics[width=0.86\textwidth]{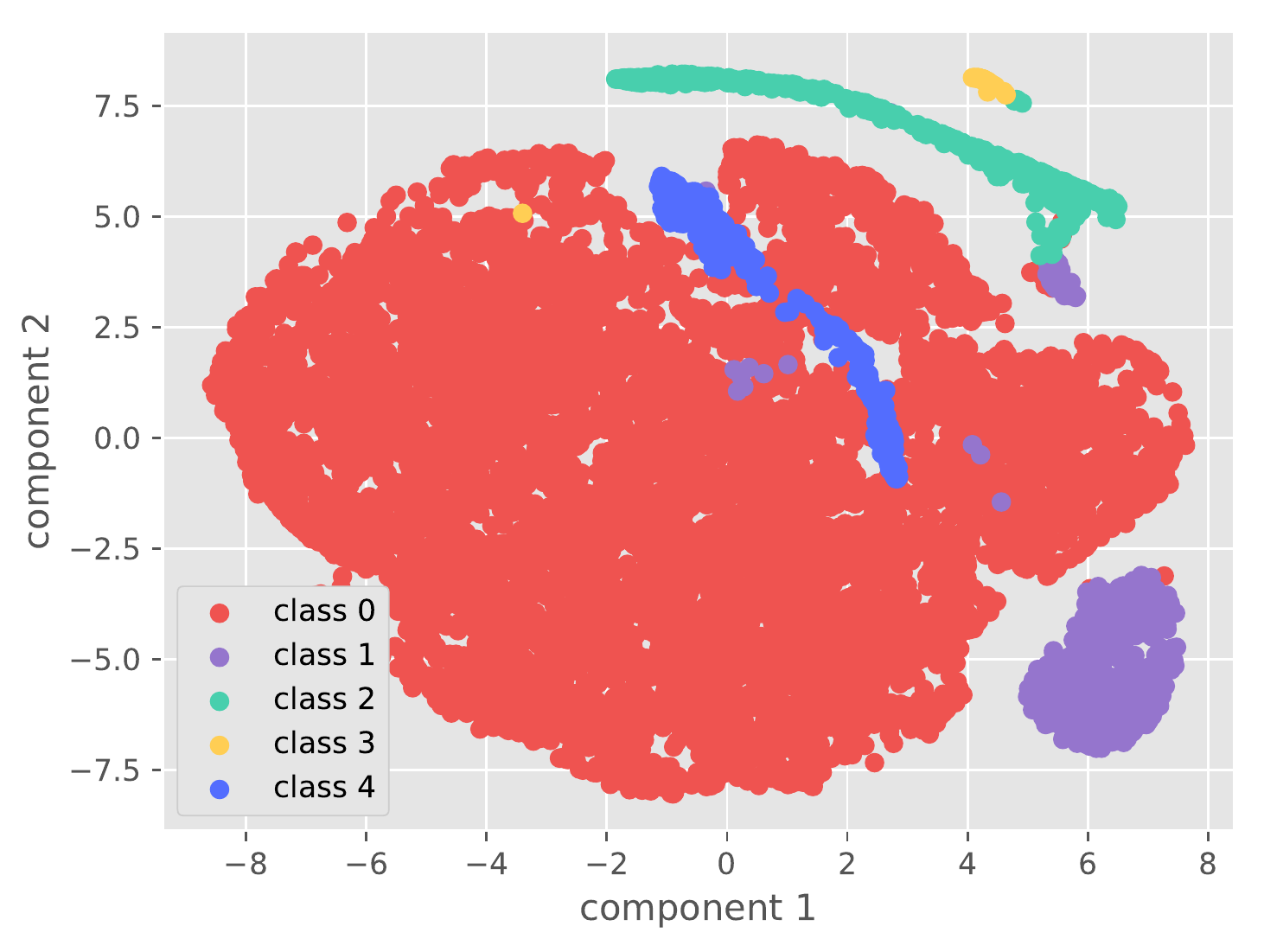}
        \caption{Learned features}
    \end{subfigure}%
    \caption{Visualization of raw data and learned representation.
    We observe that all five classes mixed together in raw data, but samples clustered well by their labels in learned features.
    Class 0-4 denote the categories of N, S, V, F, and Q, respectively.
    }
    \label{fig:visualization_analysis_5}
\end{figure}

\subsection{Convergence analysis} 
\label{sec:convergence_analysis}
\begin{figure}[t]
\centering
  \includegraphics[width=0.8\linewidth]{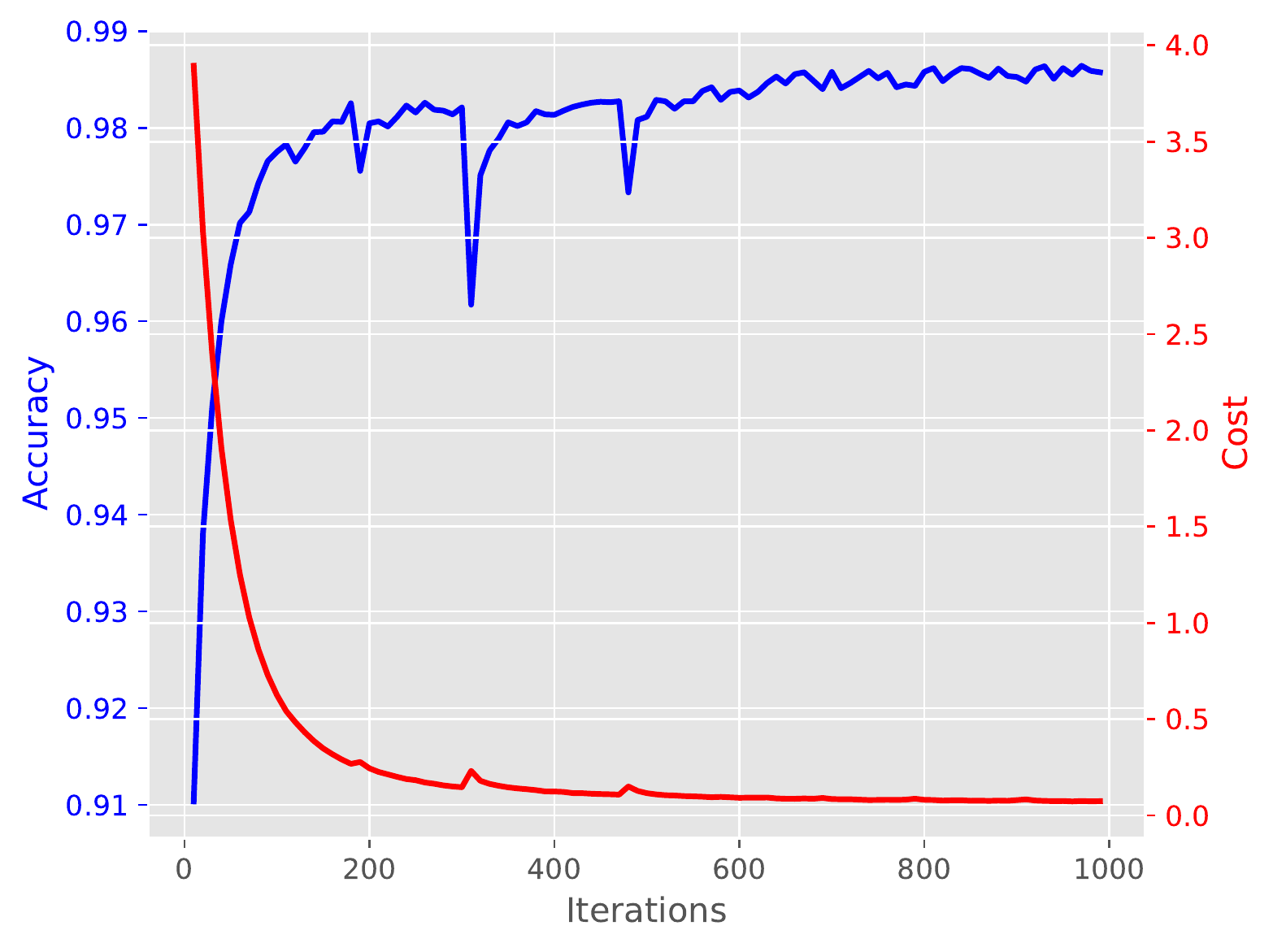}
  \caption{Accuracy and testing loss varies with the number of iterations in ABCNN.
  The highest accuracy achieves 98.57\%.
   }
  \label{fig:cost_5}
\end{figure}

Next, we evaluate the convergence characteristic of the proposed ABCNN. In specific, we check how the performance and cost varies with the training process in Figure~\ref{fig:cost_5}.
We find that the accuracy increases sharply to about 98\% in 300 iterations but then drop to 96\% near the 350-th iteration. One possible explanation is that the gradient function of neural network is not a convex optimization and the performance shifts between adjacent local optimal positions. Even though, the proposed ABCNN model can find a low cost local optimal solution, while the accuracy and cost converge to a high level
in a long training process such as 1,000 iterations.

\subsection{Visualization} 
\label{sec:visulization}    
To visually demonstrate the classification results of ABCNN for heart arrhythmia detection, we plot the visualization based on the test dataset in two parts: raw data of MIT-BIH dataset as well as the learned features of the proposed classification model. The Principle Component Analysis (PCA) is adopted to reduce the dimension of raw data to two dimension in order to visualize the data in 2D coordinate system. 
%
As shown in Figure~\ref{fig:visualization_analysis_5} (a), the raw data of 5 heart arrhythmia classes all mixed together. In contrast to raw data visualization, the learned representation in Figure~\ref{fig:visualization_analysis_5} (b) clustered well and have very clear boundaries between different classes.

\section{Discussion and Future Work} 
\label{sec:discussion}
The open challenges and future scope of the proposed approach ABCNN will be discussed in this section.
ABCNN is demonstrated effective with competitive heart arrhythmia diagnosis performance over the benchmark dataset.
However, there are still opening questions to be addressed in future work.

To begin with, the performance deep learning models highly rely on hyper-parameter tuning. We use preliminary experiments to select the best pre-defined values for hyper-parameters including learning rate, convolutional layers, kernel size, and pooling size. On the other hand, some automatic tuning methods, e.g., Orthogonal Array method~\cite{zhang2019deep}, could be utilized to save time and get better parameters. 



Second, there are still a number of potential concern  that need to be tackled before AI-based clinical diagnosis like heart arrhythmia detection to be implemented in practical scenarios. As we know, machine learning methods with great performance offline might fail to achieve good results online.
Thus, one key research direction in the future is to develop robust and stable diagnosis system.



\section{Conclusion} 
\label{sec:conclusion}
In this work, we propose a novel neural network, ABCNN, for heart arrhythmia detection. The model integrates the advantages of multi-head attention mechanism and convolutional network. 
We design an attention layer that encourages the model to focus on the most informative ECG signals.
We adopt convolutional layers to automatically capture the spatial features from the input raw EEG data.
In order to evaluate the validation of the proposed model, we conduct extensive experiments over a benchmark dataset for heart arrhythmia diagnosis. 
Our approach achieves the highest AUC for the arrhythmia diagnosis, outperforming widely used baselines. The experimental results show that ABCNN is effective and efficient in heart arrhythmia detection.

\bibliographystyle{IEEEtran}
\bibliography{pubs.bib}

\end{document}